\documentclass[10pt,conference]{IEEEtran}
\IEEEoverridecommandlockouts

\usepackage{cite}
\usepackage{lstautogobble}  
\usepackage{color}          
\usepackage{zi4}            
\usepackage{url}
\usepackage{multirow}

\usepackage{graphicx}

\usepackage{listings}
\usepackage{todonotes}
\usepackage[capitalise,noabbrev]{cleveref}
\usepackage{verbatim}
\usepackage{longtable}
\usepackage{xspace}
\usepackage{verbatim}
\usepackage{enumitem}
\usepackage{fp}
\usepackage{balance}
 \usepackage{booktabs} 

\newcommand{\tot}{28\xspace}
\newcommand{\per}[1]{$\FPeval{\x}{round(#1/28*100,1)} \x\%$}

\begin{document}

\title{Model-Driven Approaches for DevOps: \\ A Systematic Literature Review
}

\author{\IEEEauthorblockN{Hugo da Gião}
\IEEEauthorblockA{\textit{HASLab/INESC TEC} \\
\textit{Faculty of Engineering, University of Porto}\\
Portugal \\
email address or ORCID}
\and
\IEEEauthorblockN{Rui Pereira}
\IEEEauthorblockA{\textit{OutSystems}\\
Portugal \\
email address or ORCID}
\and
\IEEEauthorblockN{ Jácome Cunha}
\IEEEauthorblockA{\textit{HASLab/INESC TEC} \\
\textit{Faculty of Engineering, University of Porto}\\
Portugal \\
email address or ORCID}
}

\maketitle

\begin{abstract}
DevOps integrates tools and methodologies aimed at optimizing software development, building, deployment, and monitoring by automating and reducing the development lifecycle and improving software quality. However, despite its many advantages, it presents notable challenges, particularly in terms of usability and migration. 
On the other hand, model-driven engineering (MDE) offers a way to abstract from specific technologies and implementation details, which can be leveraged to improve these aspects, as happens in software development. 

In this paper, we present the first systematic literature review aiming to understand the current MDE approaches for supporting DevOps development. 
We focus on categorizing the works based on computing domains, phases of the DevOps process they contribute to, and the steps of reengineering they propose.

Our findings highlight a dominant focus on cloud computing, while some studies propose domain-agnostic approaches. We also observe that the deployment and integration phases are the most explored.
This shows research opportunities, both in terms of the domains and DevOps phases.

We also found limited use of text-to-model transformations, while nearly half of the papers employed model-to-model transformations. 
Text-to-model transformations were more prevalent overall. Most notably, we discovered that no study proposed a complete reengineering process, which represents an exciting avenue for future research.
\end{abstract}


\section{Introduction} \label{sec:introduction}

DevOps represents an approach to software development and delivery that employs a variety of tools and techniques to unify the traditionally separate domains of development and operations. The term originates from combining Development and Operations, reflecting a culture in which these two areas function collaboratively rather than independently \cite{DevOps}. 
This integration focuses on achieving a high degree of automation in the software development, building, deployment, and monitoring processes. The primary objectives are to ensure reliable, secure, fast, and continuous software delivery, ultimately improving business productivity and promoting employee well-being. Furthermore, software engineers who adopt DevOps practices demonstrate greater resilience to rapid changes~\cite{kim2016devops}.
This automation is achieved through the use of tools that allow software engineers to systematically describe the steps necessary to achieve their goals, from automated testing to deployment to monitoring errors in production. These descriptions are known as pipelines and are often implemented through scripts that follow the rules of the chosen tools.

Despite its advantages, adopting DevOps presents challenges for organizations and developers, including restructuring of organizations and teams, addressing issues in legacy infrastructures, and learning and integrating various DevOps tools~\cite{Ghantous2017DevOpsCP}. Furthermore, developers often struggle to learn and use specific DevOps tools, which can be daunting and may diminish the anticipated benefits of DevOps adoption~\cite{challengesdocker}.

After the initial adoption, a possible problem is migrating between different DevOps tools, as this presents several challenges due to technical, operational, and organizational complexities. Many tools rely on highly customized configurations, requiring significant effort to adapt workflows, scripts, and settings to new tools \cite{devopsmigratinchallenges1,devopsmigrationchallgens2}. 
Moreover, teams must navigate a learning curve and overcome resistance to change, which can slow productivity. Migration efforts can lead to temporary disruptions in DevOps pipelines, risking delays in releases. These challenges often demand significant time, resources, and planning to minimize downtime and ensure a smooth transition. Proper testing, phased implementation, and stakeholder engagement are critical for a successful migration~\cite{devopsmigratinchallenges1,devopsmigrationchallgens2}.
Despite the challenges, migrations are done frequently \cite{dagião2024chroniclescicddeepdive}.

These challenges are not unique to DevOps development, as they have been seen and studied in
(general) software development.
Model-driven engineering (MDE) approaches address many of these challenges, offering benefits such as streamlining the design process, generating code for diverse platforms, and enhancing communication between teams and their members~\cite{mdepractice}. By serving as partial system representations, models can abstract away unnecessary details, allowing developers to focus on the most relevant aspects at each stage. 
MDE is thus a promising approach to tackle the complexity associated with the adoption, use, and migration of DevOps~\cite{reengineering1,reengineering2,reengineering3}.
Thus, in this work, our first goal is to understand the current landscape of MDE approaches for DevOps and, in particular, to answer the following research question (RQ):

\begin{flushleft}
\noindent
\textbf{RQ1}: How well does MDE cover DevOps?
\end{flushleft}

Thus, we seek to assess the focus of existing work in terms of the coverage they provide for DevOps studying \textit{i)} the computing domains addressed by MDE approaches and, \textit{ii)} the different DevOps phases for which MDE solutions exist. By answering this RQ, we seek to find gaps in the literature that may pave future research directions, particularly by finding domains and DevOps phases not (fully) addressed by MDE approaches.
Briefly, we found that the \tot analyzed works cover a few computing domains, including cloud computing and data science, but also propose domain-agnostic solutions. Moreover, all phases are covered. However, several phases are covered only in some domains, showing significant gaps that deserve further research, as detailed in \cref{sec:rq1}.

Currently, there is little support for software engineers to migrate between different DevOps tools \cite{dagião2024chroniclescicddeepdive}.
This is one of the use cases where MDE excels. MDE reengineering approaches offer a structured method for transforming and modernizing software systems by utilizing metamodels and model transformations \cite{reengineering1,reengineering2,reengineering3}. 
A second goal of our work is to explore existing MDE approaches to fully or partially support DevOps migrations by answering our second RQ:

\begin{flushleft}
\noindent
\textbf{RQ2}: What is the MDE support for reengineering DevOps pipelines?
\end{flushleft}

We intend to gather insights about the support for the different phases of reengineering to better characterize the approaches that can be used for migration. 
Our analysis reveals that model-to-model and model-to-text transformations are common, while text-to-model and model constraints are less explored. Notably, a complete reengineering process remains unaddressed. There is still quite some space for new reengineering solutions, particularly for reverse engineering and the complete process. We present more details in \cref{sec:rq2}. 

This paper introduces the first systematic literature review (SLR) dedicated to examining MDE approaches to aid software engineers in improving their use of DevOps. Thus, we synthesize existing work at the intersection of these two fields and identify potential gaps. 



In the following section (\cref{sec:related}), we discuss related work. The SLR methodology we followed is presented in \cref{sec:methodology}. \cref{sec:quality} provides a qualitative analysis of the works used in this review. \cref{sec:threats} highlights the threats to the validity of our work.   Finally, \cref{sec:conclusions} provides concluding remarks. 

\section{Related Work}\label{sec:related}

Several literature reviews exist to individually study DevOps and MDE, but none specifically addresses the use of MDE within the context of DevOps.

Some surveys examine DevOps challenges, security practices, culture, best practices, pitfalls, and benefits. One study identifies challenges for engineers, managers, and researchers \cite{survey-1}. 
Another conducts an SLR on security challenges and practices in DevOps~\cite{survey-2}. DevOps culture is explored through an SLR too~\cite{survey-3}. 
Other works assess best practices~\cite{survey-4}, pitfalls~\cite{survey-5}, benefits~\cite{survey-7}, and its impact on software quality~\cite{survey-6}. An SLR explores the benefits and challenges of integrating Agile, DevOps, and Cloud~\cite{slrdevopsnew1}, while another examines DevOps adoption barriers in developing countries, focusing on Sri Lanka~\cite{slrdevopsnew2}. A study focuses on current and past practices in DevOps usage in open-source projects \cite{dagião2024chroniclescicddeepdive}. However, none address MDE techniques for DevOps. 

Several surveys have summarized the contributions of MDE. A study reviews 140 articles on MDE in 
cyber-physical systems (CPS), identifying research gaps~\cite{survey-12} and another explores the use of MDE to improve CPS sustainability \cite{BARISIC2025101051}. Others explore metamodels in requirements elicitation~\cite{survey-9}, MDE benefits and process reuse~\cite{survey-10}, reverse engineering~\cite{survey-11}, and cloud computing challenges~\cite{survey-13}. A mapping study examines the role of MDE in security and safety systems~\cite{survey-14}.  
SLRs have analyzed requirements engineering in model-driven development~\cite{survey-15}, integration of DevOps in MDE, highlighting automation and research directions~\cite{survey-devops-mde}, machine learning applications in MDE~\cite{slrmdenew1}, and domain-specific modeling tools and techniques~\cite{slrmdenew2}.




\section{Research Method}\label{sec:methodology}

In this section, we present our research method for conducting the SLR. In our work, we used the method proposed by Kitchenham  \cite{Kitchenham2007SR} because it provides a rigorous framework to systematically analyze and synthesize existing research, ensuring a comprehensive and unbiased review process, being a widely recognized approach for conducting SLRs. 

\subsection{Planning}

In this phase, we define the need for the review, formulate research questions, and develop a protocol that outlines the search strategy, inclusion and exclusion criteria, and data extraction methods.  

During the related work search, we have found multiple individual reviews studying MDE and DevOps, but no secondary articles studying MDE approaches for DevOps.
Moreover, during the execution of our SLR, such works were also not found. Thus, it is safe to say that there is no systematic literature review of MDE solutions for DevOps. Thus, our SLR is fundamental to understand the current situation and discover possible gaps, leading to new research paths.

The RQs and their rationale were identified in \cref{sec:introduction}.

We used the following query to search for primary studies on four major libraries, namely the ACM Digital Library \cite{acmdl}, 
IEEE Xplore \cite{ieeexplore},
ScienceDirect \cite{sciencedirect},
and SpringerLink \cite{springerlink}:

\begin{center}
    DevOps AND (``model-driven'' OR ``model driven'')
\end{center}

We performed a screening of the papers and identified several additional terms related to DevOps, such as `MLOps', `DevSecOps', `CloudOps', `AIOps', and `DataOps'. These terms represent various specialized fields of DevOps application. As a result, we decided to incorporate them into our search query.
Thus, we expanded the original query:

\begin{center}
(DevOps OR MLOps OR ChainOps OR DevSecOps OR CloudOps OR AIOps OR  DataOps)\\AND\\
(``model-driven'' OR ``model driven'')
\end{center}

Regarding the MDE part, we chose not to include terms such as model-based or similar, as our focus is on model-driven approaches, where models, metamodels, and transformations are central to the process. In contrast, model-based approaches involve models in a more lightweight manner and, thus, are not the primary focus of this study.

Based on the research questions, we wanted to identify studies that specifically applied MDE techniques across various aspects of DevOps and had the potential to contribute to the reengineering process. This selection was made using the following inclusion and exclusion criteria:

\subsubsection{Inclusion criteria}
\begin{description}
     \item[IC1]  The work applies MDE to the field of DevOps.
    We accept works that employ MDE techniques to enhance the use of DevOps. This includes publications that propose or improve techniques, tools, modeling languages, frameworks, etc., for applying MDE in any aspect of DevOps. 

 \end{description}

\subsubsection{Exclusion criteria}
\begin{description}
     \item[EC1] The work does not address the topics of DevOps and MDE.
        We reject publications whose work does not explicitly contribute to any of the fields of DevOps or MDE.
 
 \item[EC2] The paper applies DevOps techniques to MDE. 
    Several works exist that apply DevOps techniques to improve several aspects of MDE. Since those works are not the focus of this literature review, we dismiss them.
 
     \item[EC3] Secondary, tertiary, or survey works.
     These works are also dismissed since our goal is to analyze original primary works and report on the state-of-the-art of MDE and DevOps. We feature these works in related work.

    \item[EC4] (Extended) abstracts.
   We dismiss abstracts or extended abstracts since they do not provide enough information to evaluate their contributions to this field.

    \item[EC5] Proceedings and books.
    We also dismiss proceedings of conferences and books since they contain several publications, most of which may be irrelevant to our query. Relevant papers should appear in the search alone. 

    \item[EC6] Informal literature such as slides, conference reviews, and informal reports.
    The lack of scientific validation and technical details about the contributions makes these kinds of documents not suitable for our analysis.

    \item[EC7] Duplicated papers.
    Since we are using multiple libraries, it is possible that some works are indexed by more than one, and thus, we exclude such cases.

    \item[EC8] Papers written in languages other than English.
    Such works have limited accessibility and are thus excluded.

    \item[EC9] Papers with content included in other works. For instance, papers that were extended in some way, such as journal extensions or special issues, are excluded.

    \item[EC10] Papers that do not offer a specific MDE contribution in DevOps. 
    We reject works that explore how MDE techniques can enhance the state of DevOps but do not offer concrete solutions. This includes publications that gather requirements from industrial or academic case studies to apply MDE in various aspects of DevOps.

    \item[EC11] Papers that do not propose metamodels in their work.
%
      A metamodel is a central point for the reengineering process. Since one of our goals is to characterize the current support of MDE solutions for reengineering DevOps artifacts, works where the solution does not involve either a new or a reused metamodel, are excluded.
 \end{description}

Our data extraction method involved categorizing the selected papers based on their technical domain, the DevOps phases they addressed, and the MDE reengineering steps utilized. Additionally, we classified the articles according to the type of venue in which they were published and the validation methods used in their studies.

\subsection{Conducting}

In this phase, we conduct systematic searches, read and filter the studies based on predefined criteria, extract data, and assess the quality of the selected studies.  

The search was conducted in January 2025.
We applied the query to the full text of the publications, including titles, abstracts, and main content, to maximize the likelihood of identifying all relevant papers.

%

A total of 1,353 research papers were gathered from four major digital libraries. The largest portion came from SpringerLink with 902 papers, followed by the ACM Digital Library with 259 papers and ScienceDirect with 154 papers. IEEE Xplore contributed the smallest number, with 38 papers included in the collection.

\begin{figure}[!htb]
    \centering
    \includegraphics[width=.9\columnwidth]{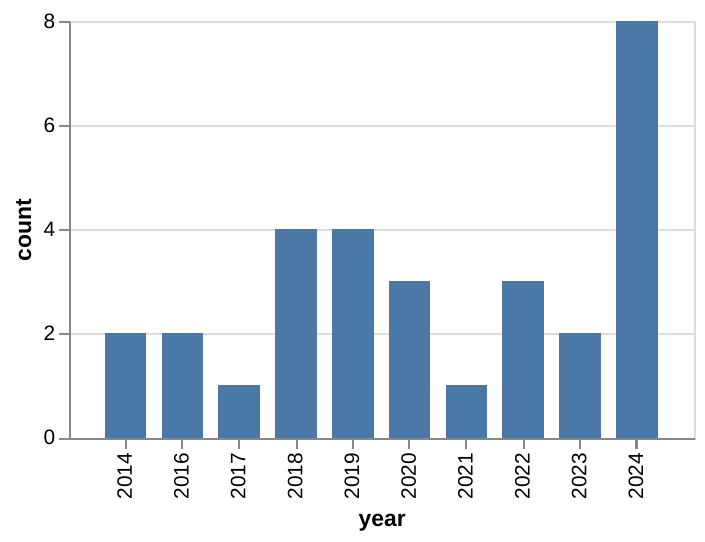}
    \caption{Publications per year.}
    \label{fig:publications}
\end{figure}

Next, we manually filtered the papers by reading their content and applying our inclusion and exclusion criteria.
Of the 1353 papers, we selected \tot publications. 
The first publication is from 2014, a couple of years after the widespread adoption of DevOps.
Notably, there was a significant increase in publications in 2024, with eight papers (\per{8}) published in that year alone, as can be seen in \cref{fig:publications}.
    
%
%

Finally, one author analyzed the \tot papers and categorized them according to our predefined data extraction method, registering the data in a spreadsheet file.

\section{How Well Does MDE Cover DevOps?}\label{sec:rq1}

In this section, we seek to answer RQ1 by understanding the current literature proposing MDE solutions for the realm of DevOps.
In particular, we intend to understand whether the proposed MDE approaches cover certain domains more than others, possibly showing the need for the community to focus their work on particular domains. Moreover, we also want to study how the works we identified address the several phases of DevOps as this may expose the phases that are not (fully) covered by MDE. Finally, we intend to explore the relationship between the DevOps phases and domains, thus assessing the current coverage of MDE under these two axes.

To answer this RQ, we formulate the following sub-research questions, which will help identify the domains, the phases, and the existing gaps at the intersection of both.

\begin{description}
    \item[RQ1.1] Which computing domains are covered by MDE?

    \item[RQ1.2] Which phases of DevOps are covered by MDE?

    \item[RQ1.3] What is the gap in the coverage of MDE to DevOps, considering its phases and computing domains?
\end{description}

\subsection{Which Computing Domains Are Covered by MDE?}\label{sec:rq11}

\begin{figure}[!hb]
    \centering
    \includegraphics[width=.9\columnwidth]{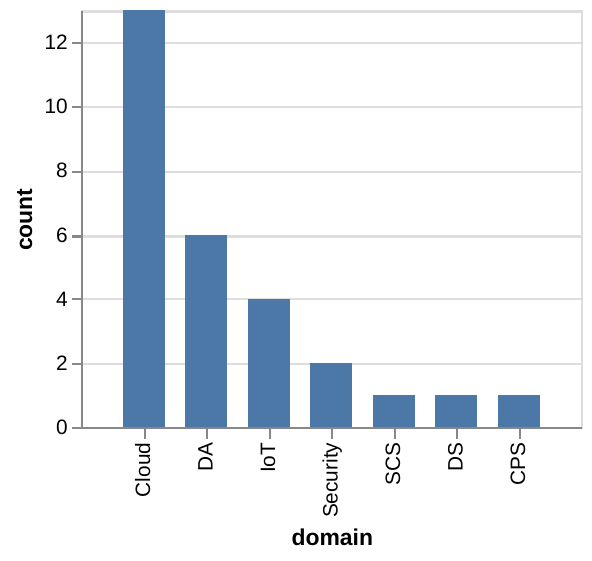}
    \caption{Domains across the papers found. DA stands for domain-agnostic, CPS for Cyber–Physical Systems, IoT for Internet of Things, and SCS for Safety-Critical Systems.}
    \label{fig:domains}
\end{figure}

To answer this question, we reviewed each of the \tot papers and identified the domains they address. This information was gathered through a detailed analysis of the papers. 
We began without predefined domains, and for each paper, we either identified a new domain or classified it under an existing one. In the following paragraphs, we describe these works, grouped by domain.
\cref{fig:domains} shows the number of works per domain.

\subsubsection{Approaches by Domain}

\paragraph{Cloud}
Of the \tot publications under study, 13 (\per{13}) focus on the cloud domain, making it the most prominent area of interest.
Seven of these studies \cite{number-24-sl,number-27-sl,number-19-sl,number-17-sl,number-15-sl,number-8-sl,standartsbasedtosca}, created their own metamodels based on TOSCA, which is a standard commonly used in the cloud computing domain~\cite{oasis2013tosca}. 
In contrast, six of them develop their own custom metamodels~\cite{number-2-sl,number-11-sl,number-27-sl,number-16-sl,apiecosystems,smartmanufacturing}, with most of these metamodels representing real-world technologies and cloud computing systems.
We also found that the generation of DevOps code was a common focus, as many of these papers aiming to solve concrete technical problems within the cloud domain. A key motivation for adopting MDE and automating processes in this domain was to save time and reduce errors associated with using specific cloud tools.

\paragraph{Internet of Things}
We identified only four publications (\per{4}) related to the Internet of Things (IoT) domain. 
One of these publications proposes a metamodel aimed at standardizing current and future IoT architectures~\cite{number-7-sl}. Another study presents a MDE approach to characterize the various elements of Industry 4.0 applications, facilitating their deployment by generating and executing configuration, orchestration, and deployment files~\cite{iot40}. A third project focuses on developing a modeling infrastructure for systematic engineering of IoT applications~\cite{number-33-sl}. The final publication leverages MDE to improve reliability, quality, security, and privacy in the DevOps process for IoT systems~\cite{number-26-sl}, with the authors extending an existing modeling language to support the deployment of IoT security and privacy mechanisms running on Edge devices. 

\paragraph{Security}
We identified only two publications (\per{2}) related to the security domain. 
One paper introduces an MDE-based IDE developed 
to support the DevSecOps lifecycle. The Eclipse-based IDE integrates various security-focused tools, offering developers a unified process~\cite{piearce}.
Another paper presents a framework to identify and mitigate security risks in citizen and low-code development by linking risks to specific actors and tools. It provides actionable steps for secure development and integrates automated guidance aligned with DevSecOps principles~\cite{securityriskmanagement}.

\paragraph{Safety-Critical Systems}
For safety-critical systems (SCS), we found only one article (\per{1})~\cite{number-4-sl}.
In this work, the authors propose a modeling framework that aligns with DevOps principles, such as testing and integration, aiming to design SCS while adhering to industry standards. The authors define their modeling language using a metamodel and emphasize that the ability to reason about and make proofs over the models was a key factor in their decision to use MDE.

\paragraph{Cyber-Physical Systems}
For cyber-physical systems (CPS), we found two publications focused on modeling the entire development lifecycle of these systems~\cite{aidort,aiarchitecture}.
The authors present a software architecture that forms the foundation of an MDE framework designed to support the development of AI-augmented solutions for Industry 4.0 in critical domains such as healthcare or autonomous vehicles. The framework integrates methods and tools for software and system engineering, along with validation techniques~\cite{aiarchitecture}.
The AIDOaRT project adopts MDE combined with AI to support the full lifecycle of CPS, from modeling, coding, testing, monitoring, and continuous development, aiming to make CPS more reliable and trustworthy~\cite{aidort}.

\paragraph{Data Science}
In the field of data science (DS), the only work found applies MDE techniques to improve the reproducibility of data science projects. The authors developed a model-driven framework that enables users to define models, verification rules, and restrictions, representing DevOps pipelines for data science tasks in a platform-independent manner by separating the modeling and operation layers~\cite{number-35-sl}.
From these, users can generate code and DevOps pipelines.
%

\paragraph{Domain-Agnostic}
We found six publications (\per{6}) whose work is not restricted to any particular domain \cite{number-12-sl,number-31-sl,automatedpipelines,interactivemodelbasedrequirements,doml,pipelinesfeelings}, which we refer to as domain-agnostic (DA).
%
%
In one of these works, developers can use the proposed framework to create and manage DevOps processes and platforms, including the sequence of processes and the overall application of service processes~\cite{number-12-sl}. 
Another project uses a metamodel based on UML deployment diagrams to automate deployment, which the authors extend with additional features. The authors aim to increase the frequency and quality of software delivery by employing MDE to automate deployment \cite{number-31-sl}. 
%
One paper proposes a multi-platform language and editor for analyzing automation pipelines across different platforms. 
The study identified five common pipeline patterns, showcasing the potential for cross-platform analysis~\cite{automatedpipelines}.
%
In another publication, the authors present an MDE approach for collaborative projects, addressing (non-)functional and data requirements, which can be seen as part of the operations phase of DevOps. They introduce an iterative requirement engineering (RE) process with microtasks and a dedicated modeling language \cite{interactivemodelbasedrequirements}. 
%
Another paper introduces the DevOps Modeling Language (DOML), an Infrastructure-as-Code (IaC) modeling language for infrastructure deployments. DOML simplifies resource provisioning and application deployment by using a single modeling paradigm that can map to multiple IaC languages, reducing the need for deep expertise in specialized IaC tools. The authors compare DOML with state-of-the-art IaC languages,
demonstrating its advantages through examples and a case study~\cite{doml}.
Finally, one publication highlights the limited effectiveness of integration and deployment pipelines (two phases of DevOps) due to the lack of justification in their structure. To address this, the authors propose to extend the Justification Diagram~\cite{justificationdiagrams} model to support iterative and incremental development of pipelines using patterns and merging syntax. This approach aims to better articulate integration and deployment pipelines into the DevOps process, resulting in more maintainable and well-justified pipelines~\cite{pipelinesfeelings}.

\subsubsection{Answering RQ1.1}

In the \tot papers analyzed, we identified six distinct computing domains and several domain-agnostic works that can potentially be applied in multiple domains. In total, 22 papers (\per{22}) focus on domain-specific approaches. 
This may happen for multiple reasons. A possible explanation is that researchers find it more effective to propose focused approaches. It can also be the case that proposing generic solutions is too challenging.

Most researchers focus their efforts on the cloud domain, accounting for almost half of all reviewed studies (\per{13}). 
The investment of researchers in the cloud domain may be motivated by the fact that the rise of cloud computing brought new problems (such as quite diverse systems), which pushed for solutions such as the adoption of DevOps, making these two concepts quite tangled \cite{7578902}.

However, many other domains remain either underrepresented. 
For instance, SCS and CPS have not been extensively covered, signaling opportunities for further research. 
Moreover, several domains, such as mobile development or high-performance computing, have not yet been explored at all. This indicates a significant gap in the literature that presents opportunities for future research endeavors.


\subsection{Which Phases of DevOps Are Covered By MDE?}\label{sec:rq12}

We now analyze the articles and categorize them according to the phases of the DevOps lifecycle they address. We follow the DevOps phases proposed in~\cite{phases}, which include development (DEV), testing (T), integration (I), deployment (D), monitoring (M), feedback (F), and operations (O). 
By organizing the literature according to the DevOps phases, our goal is to understand how MDE techniques contribute to each stage of the DevOps process and to identify any existing gaps.

\begin{figure}[!hb]
    \centering
    \includegraphics[width=.9\columnwidth]{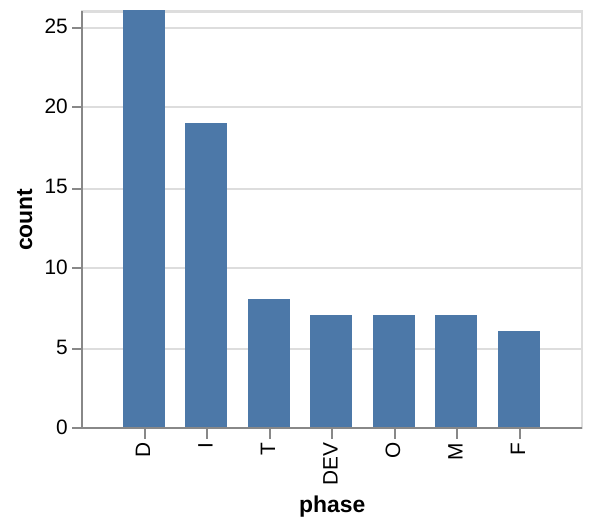}
    \caption{Number of works addressing each of the DevOps phases. We use the following abbreviations: development (DEV), testing (T), integration (I), deployment (D), monitoring (M), feedback (F), and operations (O).}
    \label{fig:phases}
\end{figure}

\cref{fig:phases} presents an overall view of the contributions made to each DevOps phase.
It is important to note that a single paper may cover multiple phases of the DevOps lifecycle. In such cases, we count the study in all relevant phases it addresses.
The most explored phase is deployment, followed by integration. These two phases are often studied together, as further discussed in this section. Other phases receive comparatively less attention in the literature.
In the following paragraphs, we discuss the papers according to the phases they cover, starting with the ones that cover all phases.

\subsubsection{Approaches by Phases}

\paragraph{All Phases of DevOps}
We identified six papers (\per{6}) that cover the entire DevOps pipeline. 

A study introduces an MDE framework for defining pipelines across conceptual and operational layers. By enabling reproducibility, replicability, and model validation, this work directly contributes to DevOps by strengthening automation and ensuring tool independence~\cite{number-35-sl}.
Another work develops an Eclipse-based IDE that integrates heterogeneous tools into a single environment. This integration advances DevOps by allowing developers to orchestrate the entire lifecycle more efficiently, reducing fragmentation and improving process traceability~\cite{piearce}.
A different study applies MDE to build a framework that maps security risks in low-code platforms using established security methods. By embedding prompts generated from models, it enhances DevSecOps practices, ensuring that security becomes a seamless part of the DevOps pipeline rather than an afterthought~\cite{securityriskmanagement}.
Another approach presents DevOpsML, an MDE-based framework that unifies processes and platforms. Its contribution to DevOps lies in simplifying complex workflows for non-technical users, particularly in low-code settings, and enabling model-driven automation across all stages of the DevOps lifecycle~\cite{number-12-sl}.
One paper introduces a language and analysis environment for automating multi-platform pipelines. This contribution strengthens DevOps by standardizing pipeline creation, ensuring cross-platform consistency, and improving the pipelines' execution speed and accuracy~\cite{automatedpipelines}.
Finally, the DevOps Modeling Language (DOML) is proposed to streamline infrastructure provisioning and application deployment. DOML supports DevOps by reducing reliance on specialized expertise and offering a unified modeling approach that translates into several IaC languages, thus increasing accessibility and portability~\cite{doml}.

\paragraph{Development, Testing, and Monitoring}
The single paper (\per{1}) addressing these phases proposes MDE to build an artificial intelligence-augmented framework for CPS development, 
using prescriptive modeling for continuous development, ensuring traceability and synchronization, and enhancing modeling and validation with artificial intelligence (AI)/machine learning (ML), especially in the automotive domain~\cite{aidort}.

\paragraph{Integration, Testing, and Deployment}
The only publication tackling these phases simultaneously enhances DevOps processes through the use of AI-augmented MDE techniques~\cite{number-4-sl}.  
The project uses tools such as EMF Views and Modelio to enable integration, while MOMOT and TWIMO optimize testing with AI/ML. This work uses DevOpsML for deployment, and tools like MDE4JSON and AutomationML facilitate efficient artifact management. Overall, this approach enhances automation and reliability in these phases.

\paragraph{Integration and Deployment}
We identified 12 articles (\per{12}) that address both phases, indicating that they are often coupled together.
Part of these publications focuses on standardizing cloud orchestration and automation tools. In particular, one project improves DevOps usage by standardizing artifacts through TOSCA~\cite{standartsbasedtosca}. Another introduces CloudCAMP, a model-driven framework that automates the orchestration of cloud services, reducing manual IaC solutions~\cite{cloudmap}. A third study proposes a programming model to simplify the implementation of TOSCA-based cloud applications, reducing complexity by automating endpoint exchanges and using code generation for deployment~\cite{zimmerman}.
Three works focus on multi-cloud orchestration and interoperability. The first proposes the Cloud Modeling Framework, a domain-specific language, which supports multi-cloud provisioning, deployment, and adaptation, addressing interoperability across cloud platforms~\cite{number-11-sl}. The second uses TOSCA for cloud resource orchestration, designing artifacts independently of DevOps tools and allowing translation between TOSCA and these tools~\cite{number-19-sl}. The third improves support for multi-cloud infrastructure, enabling seamless migration and addressing issues related to privacy and performance~\cite{number-27-sl}.
One study automates software deployment to edge devices using an MDE approach with platform-independent models and constraint solving for deployment efficiency, integrated into a DevOps toolchain for edge applications~\cite{number-16-sl}. 
Another publication integrates security methodologies into a model-driven approach, linking risks to specific actors and tools during system design~\cite{number-26-sl}. 
Edmonds et al. present a project that integrates OCCI \cite{occi} and TOSCA standards for the unified management of cloud infrastructure and applications, automating the design, deployment, and orchestration of cloud applications using~\cite{number-15-sl}.
Urano, an automation tool for software deployment, extends UML to capture configuration and update requirements, improving the efficiency and efficacy of deployments~\cite{number-31-sl}. 
Another paper presents a metamodel to standardize IoT architectures, enhancing the deployment and integration phases, and facilitating development automation and testing~\cite{number-7-sl}.
Finally, a study extends the Justification Diagram model to improve integration and deployment pipeline development. This approach enables creating more maintainable and integrated pipelines through iterative development, patterns, and merging syntax~\cite{pipelinesfeelings}.

\paragraph{Deployment}
We identified seven projects (\per{7}) that focus exclusively on the deployment phase, reflecting significant research interest in improving this phase.
%
One project improves deployment by introducing an MDE approach for defining API service licenses and automating license assessment and composition to ensure compatibility between providers and consumers. This facilitates rapid application deployment through automated license validation and dependency analysis~\cite{apiecosystems}. Another project uses MDE to automate cloud application (re-)deployment using the DICER tool, which maintains quality-aware processes for efficient DevOps workflows~\cite{number-17-sl}. Additionally, a framework standardizes DevOps artifacts using TOSCA, enabling automation and interoperability across various DevOps tools and cloud applications \cite{number-24-sl}. 
%
%
A smart manufacturing project applies MDE and a visual editor to design fog-compliant applications, enabling the creation of containerized applications ready for deployment with minimal technical expertise~\cite{smartmanufacturing}.
%
%
One study proposes a model-driven approach to automate cloud infrastructure provisioning, reducing the manual effort involved in configuration management~\cite{number-2-sl}. Similarly, MontiThings, 
enhances IoT application deployment by automating error handling and separating business logic from implementation details, improving system reliability and deployment efficiency~\cite{number-33-sl}. Another approach presents an MDE tool that automates the configuration, orchestration, and deployment of data-intensive applications, addressing real-time, security, and hardware requirements in complex environments~\cite{iot40}.

\paragraph{Operations}
We identified one article (\per{1}) that focuses on the operations phase of DevOps. This study highlights the benefits of an MDE requirements engineering approach in managing complex software solutions~\cite{interactivemodelbasedrequirements}.  
%
By providing a central model for consistency throughout the project lifecycle, it supports practical deployment and improvement.

\subsubsection{Answering RQ1.2}
We identified a large number of papers for the deployment phase, totaling 26 (\per{26}), compared to the other phases. This is primarily due to the extensive focus on improving compatibility between different cloud providers, as well as approaches that combine this phase with others in the DevOps process, especially the integration phase.  

Most of the other phases were covered as part of approaches addressing the entire pipeline or in conjunction with other phases. Deployment and operations are the only phases addressed on their own. Solutions for each phase individually would allow for a more flexible combination of approaches.


\subsection{What Is the Gap in the Coverage of MDE to DevOps, Considering Its Phases and Computing Domains?}\label{sec:rq13}
\begin{figure}[!hb]
    \centering
    \includegraphics[width=.9\columnwidth]{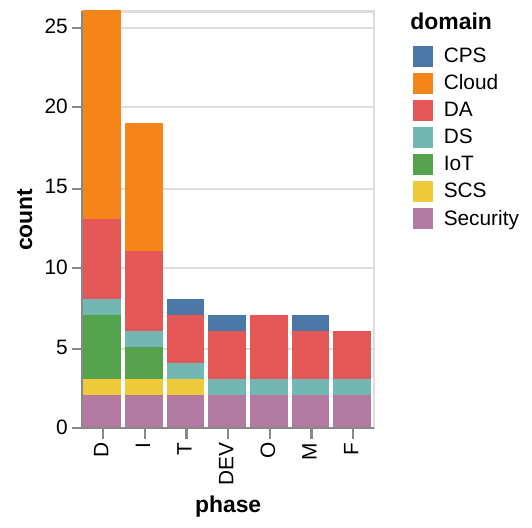}
    \caption{Number of works grouped by DevOps phase and domain.}
    \label{fig:phasedomain}
\end{figure}

In this section, we conduct an analysis that intersects the identified domains with the DevOps phases. 
Each column of \cref{fig:phasedomain} presents the number of works addressing the identified phase, grouped by the domains described in \cref{sec:rq11}.

\subsubsection{The Gap By Phases}

\paragraph{Deployment}
%
We found 26 articles (\per{26}) the deployment phase. While most of the works addressing this phase are concentrated in the cloud domain~\cite{apiecosystems,standartsbasedtosca,number-17-sl,number-24-sl,number-2-sl,cloudmap,zimmerman,number-11-sl,number-19-sl,number-27-sl,number-16-sl,number-15-sl,smartmanufacturing}, there are also studies in other domains, including 
securiy~\cite{piearce,securityriskmanagement}, 
SCS~\cite{number-4-sl}, 
IoT~\cite{number-26-sl,number-7-sl,number-33-sl,iot40},  
data science~\cite{number-35-sl}, 
as well as domain-agnostic approaches~\cite{number-31-sl,number-12-sl,automatedpipelines,doml,pipelinesfeelings}.
CPS is the only domain not covered. Nevertheless, some domains are covered by only one or two works.

\paragraph{Integration}
Our review identified 19 studies (\per{19}) that address integration in various domains. The majority of these works focus on the cloud domain, comprising eight articles~\cite{standartsbasedtosca,cloudmap,zimmerman,number-11-sl,number-19-sl,number-27-sl,number-16-sl,number-15-sl}, but all domains are covered 
(security \cite{piearce,securityriskmanagement}, 
SCS \cite{number-4-sl}, 
IoT \cite{number-26-sl,number-7-sl}, 
data science \cite{number-35-sl}), again, except CPS. Domain-agnostic solutions are also presented \cite{number-31-sl,number-12-sl,automatedpipelines,doml,pipelinesfeelings}.
The distribution in the integration phase is quite similar to the deployment one, although with a smaller number of publications.


%
%
%
%
%
%

\paragraph{Testing}
%
Eight studies (\per{8}) explore this crucial DevOps phase across five domains, namely security~\cite{piearce,securityriskmanagement}, SCS~\cite{number-4-sl}, data science~\cite{number-35-sl}, and CPS \cite{aiarchitecture}. 
Additionally, domain-agnostic proposals are also available~\cite{number-12-sl,automatedpipelines,doml}.
We now have fewer proposals covering fewer domains, too.


\paragraph{Development}
Seven studies explore the development phase (\per{7}), encompassing three domains, namely security \cite{piearce}, data science \cite{number-35-sl},  and CPS \cite{aidort}, and domain-agnostic approaches \cite{number-12-sl,automatedpipelines,doml}.  
These contributions underscore the broad applicability of MDE techniques in the development phase of DevOps. Nonetheless, research gaps persist across several computing domains.  
%
%
%
%
%


\paragraph{Operations}
%
This phase is covered by seven studies, addressing a few domains, namely security~\cite{piearce,securityriskmanagement}and data science~\cite{number-35-sl}. Moreover, domain-agnostic approaches also contribute to this phase~\cite{number-12-sl,interactivemodelbasedrequirements,automatedpipelines,doml}.

\paragraph{Monitoring}
%
Seven studies in our review examine this phase across three domains: security \cite{piearce,securityriskmanagement}, data science~\cite{number-35-sl}, and CPS~\cite{aiarchitecture}. Additionally, domain-agnostic methods also exist~\cite{number-12-sl,automatedpipelines,doml}.
On top on the few existing works, several domains have not covered in any way.


\paragraph{Feedback}
%
This phase spans only domains: security \cite{piearce,securityriskmanagement} and data science~\cite{number-35-sl}. Domain-agnostic approaches also contribute to this phase~\cite{number-12-sl,automatedpipelines,doml}. 


\subsubsection{Answering RQ1.3}

Only six papers (\per{6}) cover all phases, covering all domains~\cite{devopsml, number-35-sl, piearce, automatedpipelines, doml, securityriskmanagement}: two in the security domain\cite{piearce, securityriskmanagement},one in data science~\cite{number-35-sl}, and three domain-agnostic~\cite{devopsml,automatedpipelines, doml}. This indicates that developing a comprehensive approach across all phases is either highly complex or is not considered a viable solution.  

Moreover, several intersections remain unaddressed. For instance, the operations and feedback phases are entirely absent from work related to cloud computing, IoT, and SCS.  



The number of publications for deployment and integration is clearly more abundant compared to the other phases. This indicates that the ``Ops'' in DevOps is very much understudied, leaving plenty of interesting research paths.

Cloud computing is also only addressed in two phases (integration and deployment). Given the fact that the emergence and growth of DevOps is very much related to this domain, it is surprising that there are not more MDE approaches covering it. Once more, this shows an interesting research path.

\subsection{Answering RQ1}

Our analysis reveals that some papers concentrate on domain-agnostic approaches, but there is a significant emphasis on cloud computing. However, we observed that several computing areas remain underexplored, suggesting that applying these methods to new domains, such as mobile development, high-performance computing, or artificial intelligence, could represent a promising avenue for future research.

We identified a significant focus on integration and deployment when examining the phases covered in the studies. There is also a notable trend toward addressing all phases of the DevOps lifecycle, with increasing interest in recent years. This trend highlights the potential for leveraging these techniques in less-explored phases, such as monitoring and operations.

Regarding the intersection between domains and phases, it is notable that the focus is on integration and deployment for cloud computing. In contrast, domain-agnostic approaches tend to adopt more generic strategies that span all phases of the DevOps process. 
On the other hand, several phases are not addressed in several domains, such as development, operations, monitoring, and feedback for cloud, IoT, and SCS, or development and integration for CPS.
This indicates there are still underexplored opportunities for further investigation.


\section{What Is the MDE Support for Reengineering DevOps Pipelines?}\label{sec:rq2}

According to the literature, one should consider three phases when reengineering a system: \textit{i)} an initial reverse engineering step to construct a more abstract representation of the system under study; \textit{ii)} a step where the new representation is evolved in the direction of the new system; and \textit{iii)} a forward engineering step, that is, the creation or generation of the new system based on the representation from the previous step~\cite{10.1109/52.43044,reengineering3,reengineering1,reengineering2}.
Often, step \textit{i)} is implemented using text-to-model (T2M) transformations. The models created usually conform to one or more metamodels to enforce a correct representation. Recall that all the papers in our dataset include a metamodel. Step \textit{ii)} can be executed using model-to-model (M2M) transformations, thus evolving the system's representation into the desired direction. It is possible that these models are enriched with constraints, for instance, to impose domain or business constraints (e.g., using the Object Constraint Language (OCL) \cite{ocl}). The last step \textit{iii)} is ideally performed by model-to-text (M2T) transformations, that is, the code of the new system is automatically generated from the abstract representation.




To answer RQ2, we divide it into the following three sub-questions, according to the three phases of reengineering:

\begin{description}
    \item[RQ2.1] What is the MDE support for reverse engineering? 
    \item[RQ2.2] What is the MDE support for the evolution of DevOps models' representations?
    \item[RQ2.3] What is the MDE support for forward engineering? 
\end{description}

\begin{figure}[!hb]
    \centering
    \includegraphics[width=.9\columnwidth]{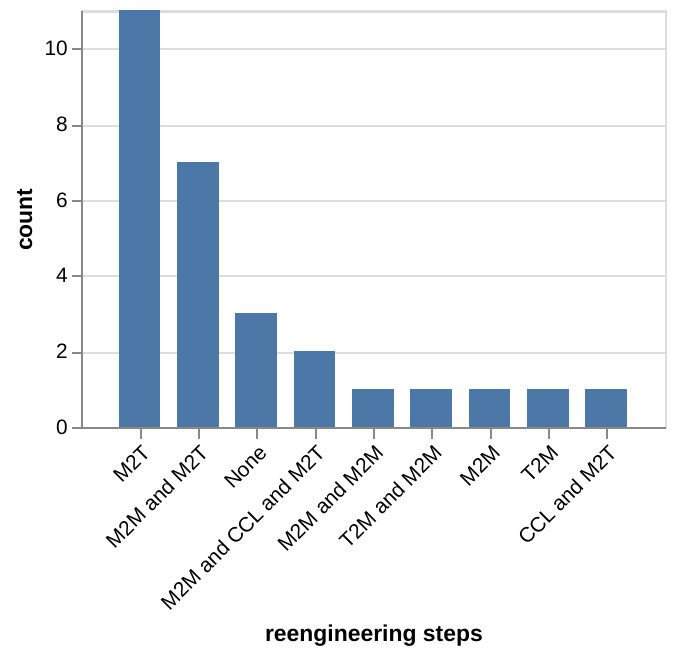}
    \caption{Number of works with contributions to the reengineering steps.}
    \label{fig:stepsreengineering}
\end{figure}

\cref{fig:stepsreengineering} presents the number of works that include T2M, M2M, and M2T transformations, as well as those that include OCL constraints. 
Note that if a paper includes multiple kinds of transformations, only the category with all kinds is increased. For instance, if a paper has T2M and M2M transformations, only the category ``T2M and M2M'' is increased, not the categories ``T2M'' and ``M2M''. In the following sections, we discuss these findings in more detail.
 
\subsection{What Is the MDE Support for Reverse Engineering?} 




We found that 26 (\per{26}) articles do not include T2M transformations.
This is the vast majority of the works, indicating that the community provides almost no support for reverse engineering existing pipelines, thus hindering their migration.

In the two cases presenting T2M transformations, the authors' goal was to parse and analyze existing artifacts addressing the phases of integration and deployment \cite{standartsbasedtosca,automatedpipelines}.

One article provides a transformation framework that converts DevOps artifacts into TOSCA-compliant models. It provides detailed examples of how specific artifacts are transformed into TOSCA node and relationship types \cite{standartsbasedtosca}.

The other article uses T2M transformations to create a multi-platform domain-specific language for DevOps pipelines, enabling the analysis and standardization of pipeline specifications across different platforms. The transformation process involves parsing YAML-based pipeline specifications, validating them against a metamodel, and generating model instances for further analysis \cite{automatedpipelines}. 

\subsubsection{Answering RQ2.1}


Both works that employ T2M transformations use this approach to validate existing DevOps scripts, covering only the integration and deployment phases.
This indicates a clear gap in the current approaches, as most do not provide any support for the initial step of a migration process. Thus, anyone trying to migrate their existing pipelines using the current MDE solutions will need to recreate them by hand as models in the current approach, which is very discouraging. Even those that include T2M transformations cover only two phases of the entire DevOps process. This paves a significant path for future research.

\subsection{What Is the MDE Support for the Evolution of DevOps Models' Representations?}


\subsubsection{Model-to-Model Transformations}
We found that most articles, 17 (\per{17}), do not propose M2M transformations. 
Of the 11 (\per{11}) articles that employ M2M transformations, several use transformations between platform-specific metamodels and platform-independent metamodels. 
This trend is evident across multiple domains. 

In the cloud domain, three studies focus on transforming platform-specific metamodels, representing various cloud tools, into TOSCA-conforming metamodels~\cite{standartsbasedtosca,number-17-sl,number-15-sl}. Three other transform platform-specific metamodels into their custom platform-independent metamodels~\cite{number-27-sl,number-16-sl,smartmanufacturing}.

In CPS, the goal is to integrate MDE with AI to create more reliable and trustworthy CPS. The AIDOaRt project applies M2M transformations to support CPS development, combining model synchronization, optimization, and automation. It also explores AI-driven modeling and uses standardized formats for data exchange, ensuring that different tools and models can work together more effectively~\cite{aiarchitecture}.  

In the single work with M2M transformations that is domain-agnostic, transformations are used to convert platform and tool models to more general approaches. M2M transformations facilitate the conversion of models across different abstraction levels, with a focus on the integration process and platform models through model weaving~\cite{number-12-sl}.

In the data science domain, M2M transformations enable the conversion between various data science models.
Their approach divides pipeline definitions into two abstraction levels: conceptual and operational models. M2M transformations are then used to map between these levels~\cite{number-35-sl}.

In IoT, M2M transformations help transition from platform-specific to platform-independent metamodels. Their method leverages M2M transformations to automate tasks like port forwarding, deployment planning, and the validation of pre- and post-conditions. These transformations refine high-level models into detailed ones, ensuring consistency in deployment.
M2M transformations also bridge the gap between business logic and implementation details, converting models into standardized formats for effective communication~\cite{number-33-sl}.

In the security domain, M2M transformations are used to convert platform-dependent models into platform-independent models. This work involves converting high-level DOML models into executable IaC languages such as Terraform and Ansible. These transformations support the DevSecOps lifecycle by linking abstract models to concrete configurations~\cite{piearce}.

\subsubsection{OCL-Like Constraints}

Only two articles employ OCL constraints in their work (\per{2}), and none use OCL alternatives. Although most works do not utilize OCL or alike, this does not imply that such constraints are not valuable. The integration of OCL constraints into an existing metamodel often requires additional tooling and expertise, which may have presented a barrier that made it less appealing. 
It may also be the case that they were simply not necessary.

In one article (in the DS domain), OCL is used to ensure the correctness and consistency of data science pipelines in their conceptual model. Specifically, OCL constraints are used to restrict the types of input and output ports in data processing elements, ensuring only valid data types are connected. OCL is also used to verify the pipeline at the model level and detect potential issues before execution. Additionally, OCL constraints are used to verify that data transformations remain consistent across different execution environments \cite{number-35-sl}. 

In an article on the IoT domain, the authors use OCL constraints in MontiThings to specify pre- and post-conditions for validating component inputs and outputs, ensuring data consistency and correctness. These constraints help to detect errors early, such as invalid sensor values, and handle them through catch statements. By separating error handling from business logic, the approach improves model reliability and maintainability, while formal verification at the model level ensures technology-independent validation~\cite{number-33-sl}.

\subsubsection{Answering RQ2.2}

Eleven works employ M2M transformations, representing less than half of all articles. They are primarily used to convert between platform-specific metamodels, representing real-world tools or platforms, and more abstract, platform-independent metamodels. Without M2M transformations, the migration of DevOps artifacts becomes severely compromised, as their evolution at the model level is not possible. We thus show the necessity of further research to complete existing approaches or propose new ones that can effectively be used for a complete migration process. 

Regarding OCL-like constraints, we found that their use is limited, with only two works explicitly using them. It is possible that their application was not clearly stated in some papers, but the overall scarcity suggests that their potential remains underexplored. This highlights an opportunity for further research, potentially by enhancing existing modeling tools to simplify the creation and integration of OCL expressions, thereby improving the overall modeling process.

\subsection{What Is the MDE Support for Forward Engineering?} 

%
%


We found that 22 (\per{22}) articles propose M2T transformations, representing the vast majority of the works explored in this review. We observed that the generated text represents real-world integration and deployment scripts for various technologies, a trend noted across multiple computing domains.

In the cloud domain, we found several uses of M2T transformations in 12 papers (\per{12}). The primary uses of these transformations are focused on generating cloud infrastructure for different technologies or creating configuration files for various cloud uses and environments. 
This includes generating REST API licenses~\cite{apiecosystems}, TOSCA YAML blueprints~\cite{number-17-sl}, configuration files for Chef cookbooks and Juju charms~\cite{number-24-sl}, provisioning of cloud infrastructure through configuration files~\cite{number-2-sl}, deriving concrete deployable models from partial models~\cite{cloudmap}, and generating configuration files for orchestrating services~\cite{zimmerman}. Additionally, these transformations are used to create deployment plans and to apply constraints during model refinement to specify compatible tools and languages~\cite{number-11-sl}, generate YAML representations for Docker Compose~\cite{number-19-sl,smartmanufacturing}, produce cloud infrastructure scripts using Acceleo templates~\cite{number-27-sl}, develop configuration scripts for deploying edge computing applications~\cite{number-16-sl} and define configurations for target environments~\cite{number-15-sl}.


In the context of SCS, we found one work focused on generating contracts for verifying system behavior~\cite{number-4-sl}.

For IoT, works focus on generating YAML files for secure IoT deployment~\cite{number-26-sl}, producing code for deploying IoT applications~\cite{number-33-sl}, and creating various elements to simplify application configuration and deployment, including generating configuration, orchestration, and deployment files \cite{iot40}. 

In the data science domain, the work found aimed at the generation of data science pipeline tasks~\cite{number-35-sl}. 

For security, we identified efforts to develop a DOML compiler capable of generating secure and executable IaC~\cite{piearce}. 

The only CPS work focuses on generating models that optimize specific objectives while satisfying given constraints~\cite{aiarchitecture}.

We identified domain-agnostic approaches aimed at generating Amelia-based executable deployment specifications~\cite{number-31-sl}, creating IaC from DOML models~\cite{doml}, and using justification diagrams to develop justified integration and deployment pipelines in GitHub Actions~\cite{pipelinesfeelings}.

\subsubsection{Answering RQ2.3}

We found that most articles propose M2T transformations, often to generate improved DevOps artifacts.
All domains include at least one work with M2T transformation, with cloud being the most represented case. However, we found a single work for most domains, which is very likely insufficient to cover all problems.

We observed that this phase of the reengineering process is the most well-supported among all three phases. Nevertheless, several articles did not include M2T transformations.
Without automatic DevOps scripts' generation, there is no true MDE. Thus, there are still some opportunities for research.

\subsection{Answering RQ2}

We observed a limited use of T2M transformations, with only two papers in our dataset employing this approach. In contrast, M2M transformations were significantly more prevalent, with about half of the studies using them. Additionally, only two articles incorporate OCL constraints in their work. We also noted that most of the articles use M2T transformations. These findings suggest promising research opportunities in the less-explored steps of reengineering. 
Furthermore, we found that not a single study fully implements the entire reengineering process. This gap highlights another potential pathway for future research in this area.

\section{Data Quality}\label{sec:quality}

In this section, we provide a qualitative assessment of the papers based on two key criteria: \textit{i)} the type of publication and \textit{ii)} the validation methods employed in the research \cite{Kitchenham2007SR}. 

Of the \tot papers reviewed, eight are workshop (\per{8}), 13 are conference (\per{13}), and eight are journal papers (\per{8}). This distribution suggests that most works are well-polished, represent complete and mature research findings, and have undergone a rigorous peer review process prior to publication.

Regarding validation, we found that four papers lack validation (\per{4}), nine (\per{9}) use an example, 13 (\per{13}) a case study, one automated evaluation (\per{1}), and another a user study (\per{1}). Although evaluation was present in most of the works, it could benefit from further improvement.

Based on this qualitative evaluation of the works, we can say they represent the current state of the art regarding MDE approaches for DevOps, and thus, our results are relevant.

\section{Threats to Validity}\label{sec:threats}

In this section, we describe possible threats to the validity of our work and how we have addressed them \cite{experimentationsoftwareengineering}.

There is the risk of missing some pertinent works.
To minimize this possibility, we designed our query to search the full text of publications, including their citations. This approach ensures that a publication discussing DevOps and MDE would need to completely omit relevant terms to be excluded, which is highly unlikely. In addition, we searched through four major digital libraries in computer science. Some libraries, such as ACM, also index publications from other publishers, further increasing the breadth of our search.
We have also augmented our initial query after a first screening of the papers, as we found some papers use other terms.

Another potential threat lies in the cases where publications do not explicitly mention their domains, DevOps phases, or reengineering steps, leading to misclassification. To mitigate this, we carefully reviewed the articles and inferred the phases and domains where they were not explicitly stated. This task was performed consistently by the same author to maintain uniformity in the interpretation of the criteria.


Finally, applying inclusion and exclusion criteria is subjective. To minimize this risk, we clearly defined the criteria, as detailed in \cref{sec:methodology}. Moreover, the same author consistently applied them, ensuring a consistent and unbiased approach.

\section{Concluding Remarks and Future Work}\label{sec:conclusions}

We identified several trends, such as the predominance of research in the cloud domain and the adoption of domain-agnostic approaches. 
However, many domains remain unaddressed, presenting valuable research opportunities. 
We found that the deployment and integration phases are the most studied, especially in the context of cloud computing. Moreover, several studies address all phases of the DevOps process using domain-agnostic strategies.
Nevertheless, there is a significant gap in several domains and phases, especially when considering the intersection of the two axes. 
This shows some possible future research paths for the MDE and DevOps communities.

Regarding reengineering, we observed a low frequency of T2M transformations, with nearly half of the papers concentrating on M2M transformations, and T2M being predominant. The use of OCL constraints was relatively rare.
Once more, there are significant gaps in the different steps of reengineering that can be explored by researchers.
One of the key takeaways from our review is the complete lack of research exploring a complete roundtrip reengineering process, which represents an exciting opportunity for future investigation.

Since multiple domains, phases, and reengineering steps are not addressed, we plan to understand whether existing approaches can be used together, i.e., to fully cover a domain or all the DevOps phases. We will also study the intersection of these approaches in terms of the metamodels they present to understand the connection between the existing solutions.


\section{Data Availability}
Our data is available in a reproducibility package~\cite{anonymous_2025_14832956}.

\balance
\bibliographystyle{IEEEtran}
\bibliography{refs_new}

\end{document}